\documentclass{elsart}

\usepackage{amssymb}

\newcommand {\um}[1]     {\; \mathrm{#1}}

\newcommand {\Cel}       {\um{\!\!\!\mbox{ }^{\circ}\mathrm{C}}}

 \newif\ifpdf
 \ifx\pdfoutput\undefined
   \pdffalse        
 \else
   \pdfoutput=1     
   \pdftrue
 \fi
 \ifpdf
   \usepackage[pdftex]{graphicx}
   \DeclareGraphicsExtensions{.pdf}
   \graphicspath{{figs/}}
 \else
   \usepackage[dvips]{graphicx}
   \DeclareGraphicsExtensions{.eps,.ps}
   \graphicspath{{figs/}}
 \fi

\begin{document}

\begin{frontmatter}



\title{Design of a Base-Board \\ for arrays of closely-packed \\ Multi-Anode Photo-Multipliers}


\author{M. Ameri},
\author{S. Cuneo},
\author{P. Musico},
\author{M. Pallavicini},
\author{A. Petrolini\corauthref{cor1}},
\author{F. Siccardi} and
\author{A. Thea}.
\address{
Dipartimento di Fisica dell'Universit\`a di Genova and 
INFN Sezione di Genova,
Via Dodecaneso 33, I-16146, Genova, Italia.
}

\corauth[cor1]{Corresponding Author: \mbox{e-mail: Alessandro.Petrolini@ge.infn.it} }

\begin{abstract}

We describe the design of a Base-Board to house Multi-Anode Photo-Multipliers for use
in large-area arrays of light sensors. 
The goals, the design, the results of tests on the prototypes and future developments
are presented.

\end{abstract}

\begin{keyword}
Photon Detectors \sep 
Multi-Anode Photo-Multipliers \sep
Housing \sep
Sensor Array \sep
High-Energy Physics \sep
Astro-Particle Physics.

\PACS 
85.60 \sep 29.40 \sep 96.40 \sep 13.85
\end{keyword}
\end{frontmatter}

\clearpage

\section{Introduction}
\label{pa:intro}

Contemporary experiments in Nuclear, Particle and Astro-Particle physics often require
the use of fast, large area and highly pixelized single photon detectors
with limited dead-areas, featuring up to hundreds of thousands of channels with tight resource limitations.
Among others, Ring Imaging Cherenkov detectors (RICH) 
(see HERAb~\cite{bi:HERAb}, LHCb~\cite{bi:LHCb} and AMS~\cite{bi:AMS}) 
and Space Telescopes for Ultra High energy
Cosmic Rays observation (see EUSO~\cite{bi:EUSO}, TUS/KLYPVE~\cite{bi:russi} 
and OWL~\cite{bi:OWL}) belong to this category.
Moreover Medical Imaging applications are often subject to very similar requirements.
Multi-Anode Photo-Multipliers are often the preferred sensor for these applications.

In such cases one of the main problems is to assemble the whole detector, on the focal surface of the
collecting optics, by closely packing 
the array of sensors, in such a way to keep an acceptable geometrical acceptance and avoid
defocusing effects.
With this goal in mind we have carried on the design, prototyping and testing of the housing 
(Base-Board) for a large array of closely-packed Multi-Anode Photo-Multipliers, with particular
emphasis on possible Space applications. As Space applications have a number of tight additional
requirements with respect to ordinary applications, the device described in this paper might have a
broad range of applicability.

One should note that the geometrical acceptance of the sensor itself is a closely related issue,
affecting the overall geometrical acceptance of the array of sensors.
The usual way to deal with the geometrical acceptance of the sensor 
consists in either improving the geometrical acceptance of the bare sensor
(like, for instance, in the R8520 and Flat Panel PMT series from Hamamatsu
Corporation~\cite{bi:NewPMT}) or using a
suitable Light Collection System (either an imaging system or not) in front of the 
sensor~\cite{bi:MyPap,bi:EUSO-LCS-notes-JP,bi:EUSO-LCS-notes-FI}.
This paper will not be concerned with this issue, and it will be assumed that the geometrical
acceptance of the sensor itself has been already maximised in a suitable way.
However, in the development of the Base-Board, we kept in mind that each sensor might be equipped
with its own suitable Light Collection System.

The Base-Board which is the object of this paper was first developed in the framework of the
development of the RICH detectors of the LHCb experiment~\cite{bi:lhcb-paper}. 
Later on the design was improved to adapt it to the Space
requirements of the EUSO mission~\cite{bi:mech}.

\section{The Goals of the Design}
\label{pa:goals}

A modular architecture is preferred when building large detectors made by assembling smaller units,
because it has many advantages including: independence of the different modules, fault propagation
limitation, easier spare modules management and procurement.  Moreover a modular architecture makes
the development, design, integration and testing phases easier.

Therefore the full apparatus shall consist of 
small independent functional units, named Elementary-Cells (EC),
assembled into larger modules, named Photo-Detector Modules (PDM).  
The Elementary-Cell shall consist of a limited number of
MAPMT, sharing some common resources and constituting a totally autonomous system.

The sharing of resources improves the economy and makes design, production and testing easier.
Moreover when resources are limited, as it is the case in Space applications, saving on resources
can be accomplished by combining functions, which is the approach implemented in the current design of the
Base-Board.

The Elementary-Cell was conceived with the following guidelines and requirements in mind.
\begin{itemize}
\item
The EC shall be an autonomous and 
compact module with different functionalities integrated
onto a Printed Circuit Board (PCB) Base-Board. 
\item
To EC shall allow to pack as closely as possible the MAPMT on the Base-Board 
and shall be built to allow to pack as closely as possible different EC, in both cases ensuring a precise
relative positioning.
The packing of both the sensors in the EC and the EC in the PDM has to be optimised to reduce losses in the
geometrical acceptance, due to dead regions between the close packed
elements, and defocusing effects, originating from a positioning of the
elements at some distance from the ideal focal surface.
\item
To EC shall allow a sharing of different resources between different MAPMT:
\begin{itemize}
\item
there is one common mechanical supporting structure and common thermal dissipation capabilities; 
\item
the thick PCB is used
as a mechanical supporting structure but it also acts as an electric board housing the electrical
components and connections;
\item
voltage dividers and HV/LV power connections may be common to different MAPMT (whenever this is desired);
\item
common electro-thermo-mechanical components such as cables, connectors and heat bridges;
\item
possibly common front-end electronic chips for more than one MAPMT.
\end{itemize}
\item
The EC shall include the front-end electronics chip for local processing of the signals, located as
close as possible to the MAPMT in order to minimise the length of the connections to preserve the
fast signal from the MAPMT, with good signal to noise ratio. 
Due to the large number of channels and the tight space, speed and power requirements
the development of an ASIC device is mandatory (see~\cite{bi:ASIC}).
\item
Finally the EC shall be a single, self-contained and autonomous system, designed as a
general purpose instrument to be used in other applications too.
\end{itemize}

The EC can be built as a thick multi-layered Printed Circuit Boards
(PCB). A number of these modules, each one making an essentially
autonomous system, are then put together to make a PDM. These are
independent assemblies tied to each other by a common supporting
structure and having a shape determined by the layout of the focal
surface. In fact a modular structure made of small elements is well suited to 
fit any focal surface shape, as it is required by some of the applications, featuring a highly curved
focal surface.

In the current work
MAPMT are  assumed to be packed into a $(2\times 2)$ EC. 
A realistic (and possibly conservative) MAPMT pitch is assumed to be
$27.0$ mm to account also for the large tolerances on the MAPMT
dimensions provided by the manufacturer: $ ( 25.7 \pm 0.5 ) $ mm.

It will be assumed that a PDM is made of an array of close-packed
EC with a suitable layout and shape, possibly surrounded by a border of
variable thickness running all around the MAPMT, to leave space for
the mechanical assembly of the PDM. Obviously the detailed design of the
PDM geometry depends on the specific application and it will not be discussed in this paper.

\section{Design, production and assembly of the Base-Board}
\label{pa:design}

The MAPMT used in this design is the R7600-M64 Multi-Anode Photo-Multipliers from Hamamatsu
Corporation~\cite{bi:Hama}.
The sensor has a square input window with $L=25.7$ mm side and about $35$ mm height. Its mass is
about $30$ g.

The current prototype has been designed and manufactured without the front-end chip, due to the
parallel development of the ASIC chip itself. In place of the front-end chip two high-density
connectors per MAPMT, with $0.5 \um{mm}$ pitch, were installed,
The signals are then taken away from the EC by means of a flex-cable to be processed by a suitable
electronics.

\subsection{The Elementary-Cell}

The EC, in the current design and configuration, consists of:
\begin{itemize}
\item
the sensors (currently four R7600-M64 Multi-Anode Photo-Multipliers);
\item
the Light Collector System;
\item
the HV dividers (currently one for each MAPMT);
\item
the front-end electronics chip;
\item
the connectors for HV/LV, signals and controls;
\item
the Base-Board, a thick PCB housing all the other components and carrying the electrical connections;
\item
a copper layer, buried inside the PCB, to help heat transfers away from the EC;
\item
mechanical elements; 
\item
a potting resin;
\item
any other required structural or functional element (if any).
\end{itemize}


A resistive voltage divider was chosen for maximum reliability.
MAPMT are powered in negative polarity, that is with anode grounded, in order to avoid using
decoupling capacitors at the anodes, thus saving space and increasing the reliability.

In order to keep the linearity of the response,
under the expected photon rates, suitable capacitors were inserted in parallel
to the last three dynodes.
Moreover, in order to save power while keeping an acceptable linearity of the response,
three different HV inputs were provided to the MAPMT at the level of the cathode and the last two
dynodes.

\subsubsection{The Base-Board}

Any EC is totally autonomous and it is individually fixed
to the PD module supporting back-planes. This approach allows a large
flexibility in the choice of the shape of the PD module, including,
possibly, allowance for a curved shape.

The Base-Board acts as the structural element,
housing the MAPMT, the front-end electronics and all the other necessary
components. It includes the traces to carry the signals from the MAPMT
to the front-end electronics and to the outside of the Base-Board as well as the traces carrying
the power and control signals to the components.

The MAPMT side of the Base-Board also houses the components of the
voltage dividers to power the dynode chain. Up to one voltage divider
per MAPMT (that is four in total) can be housed onto the Base-Board (as
it is in the prototype). Alternatively, one might install only one or
two voltage dividers (trade-off).
In the current base-line one voltage divider per MAPMT is used in order to decouple as much as
possible the four MAPMT. This approach increases the reliability and reduces the interactions among the MAPMT,
while testing the design in its most challenging configuration.

\subsection{The Mechanical and Thermal design}

The following assumptions have been made:
\begin{itemize}
\item
standard printed circuit board manufacturing accuracy ($0.1 \um{mm}$);
\item
maximum MAPMT side length limited to $25.7 \um{mm}$;
\item
minimum installation clearance between neighbouring MAPMT: $0.5 \um{mm}$.
\end{itemize}

\begin{figure}[ht]
 \begin{center}
  \fbox{\includegraphics[width=0.90\textwidth]{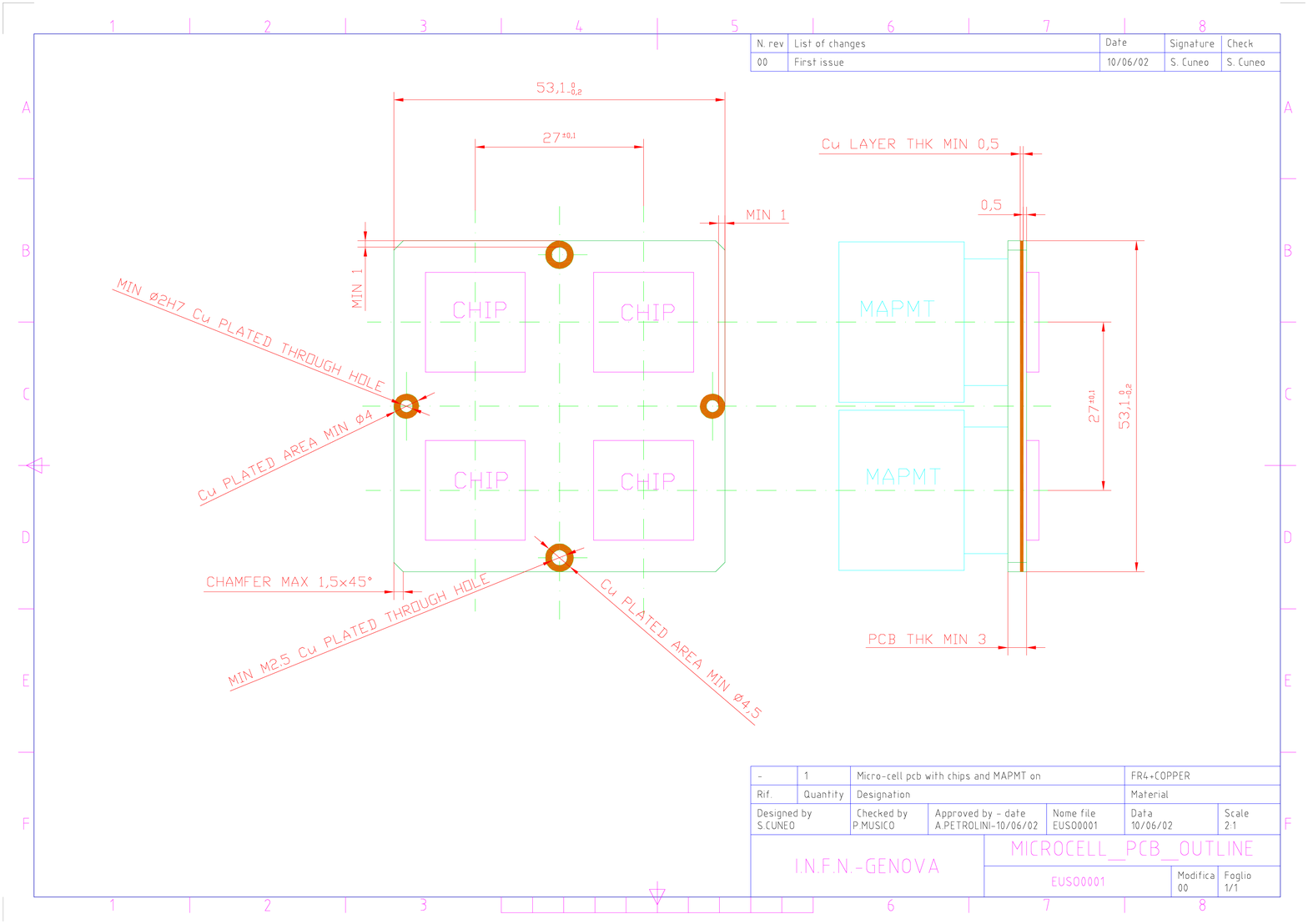}}
  \caption{Mechanical drawing of the current Base-Board design.}
  \label{fi:bbmech}
 \end{center}
\end{figure}

The mechanical drawing of the current Base-Board design is shown in figure~\ref{fi:bbmech}.

The Base-Board is a $4 \um{mm}$ thick flame retardant FR4 glass epoxy laminate.  It integrates a
copper layer, $0.5 \um{mm}$ thick, to help to convey the heat produced on the Base-Board to the
screws and dowel pins as well as improving the structural properties.  The Base-Board is secured to
the back-standing support by four $M2.5$ stainless steel screws and, if necessary, by up to four 
$\phi 2.0$ dowel pins.  This allows a precise relative positioning of the different EC
onto the PD-module and redundant support.

Screws and dowel pins fit into metallized holes drilled into the PCB.  
Screws are equipped with
cylindrical hollow aluminium spacers, so that enough space to allow the routing of the services is
available between the Base-Board and the back-standing support.  
The spacers also allow to improve
the structural performance of the fastening system.  
The location of the holes is a trade-off between the structural requirements
and the electrical layout.  
High accuracy, conventionally available in PCB
manufacturing, should allow the precise positioning of the EC, so that the design
characteristics can be kept without any interference during assembly and dismounting operation.

Screws, dowel pins and spacers also act as thermal bridges to convey the heat generated on the
Base-Board, by the voltage dividers and front-end electronics, away from it, to the back-plane
supporting structure.  
This is accomplished by conduction, and dedicated
heat bridges, whenever necessary.

\subsubsection{Structural analysis}

A finite elements structural simulation has been performed, modelling the Base-Board with its
links to the back-standing structure and simulating the MAPMT as added masses.  
The modal analysis of
a $5 \um{mm}$ thick Base-Board, fastened through four $M2.5 \times 15$ 
stainless steel screws, with an overall mass of
$0.2 \um{kg}$, has determined the following first modal frequencies: 
$716 \um{Hz}$, $872 \um{Hz}$, $1259 \um{Hz}$, $1801 \um{Hz}$, $1878 \um{Hz}$.
Afterwards a
random vibration analysis has been performed with the input spectrum suggested by NASA for acceptance of
items whose mass at launch is below 23 Kg (\cite{bi:NASA-specs}).  Such spectrum of acceleration is
equivalent to $6.9 \um{G}$ RMS. 
A standard damping ratio of $0.025$ has been assumed, constant throughout the
frequency range. The calculated response of the system, when the excitation is parallel to the
Base-Board plane, showed a sharp peak
in correspondence of the first modal frequency, with an integral value of $25 \um{G}$ RMS. 
The calculated response of the system, when the excitation is orthogonal to the Base-Board
plane, showed a sharp peak in correspondence of the fifth modal frequency, with an integral value of
$39 \um{G}$ RMS. 
The result is consistent with the fifth modal shape.  

The static
analysis performed with the above acceleration combined in a load vector showed the screws working
as fully constrained beams, as expected, with the most stressed points at their ends. Calculated Von
Mises' equivalent stress in those points was well below the allowed limits.

\subsubsection{Thermal analysis}

A finite elements analysis has been performed on a simplified model, where the EC was
thermally linked, through its fastenings, to a cold pit at $T=13\Cel$.  Heat is supposed to be
drained away by means of conduction only.  An uniformly distributed power of $0.3 \um{W}$, as
expected in real operation, has been assumed.  The results of the analysis showed that the copper
layer works well: the PCB hot side temperature becomes uniform to within one degree and a temperature drop of
$\Delta T=6\Cel$ is generated with respect to the cold pit.

\subsection{The Electrical design}

The routing of the electric connections is hard, due to the high number of channels and components
and the limited available space.  However, thanks to the regular geometry of the MAPMT anode
connections and of the connectors, it was easy to manually route the connections in such a way that
no intermediate layer was necessary in the PCB for the routing the signal paths.

The four voltage dividers (made of resistors, capacitors and the three HV inputs per MAPMT) are made
of surface-mount devices and installed on the side of the MAPMT.

The signal connections starts from the MAPMT side and are routed to the back-side of the Base-Board
through via holes, where either a front-end chip or suitable connectors are housed.

The signal connectors on the back-side are located at the centre of the corresponding MAPMT.  The HV
connector is located at the centre of the Base-Board and the three voltages are routed to the four
voltage dividers.

The PCB layout was done following the standard design rules.

\subsubsection{Mounting of the MAPMT}

By using a surface-mount technique for the MAPMT the thickness of the Base-Board can be kept
relatively small, compatibly with the mechanical requirements, thus saving mass.  The optimal
thickness will result from a trade-off with the structural requirements.

\begin{figure}[htb]
 \begin{center}
  \fbox{\includegraphics[width=0.75\textwidth]{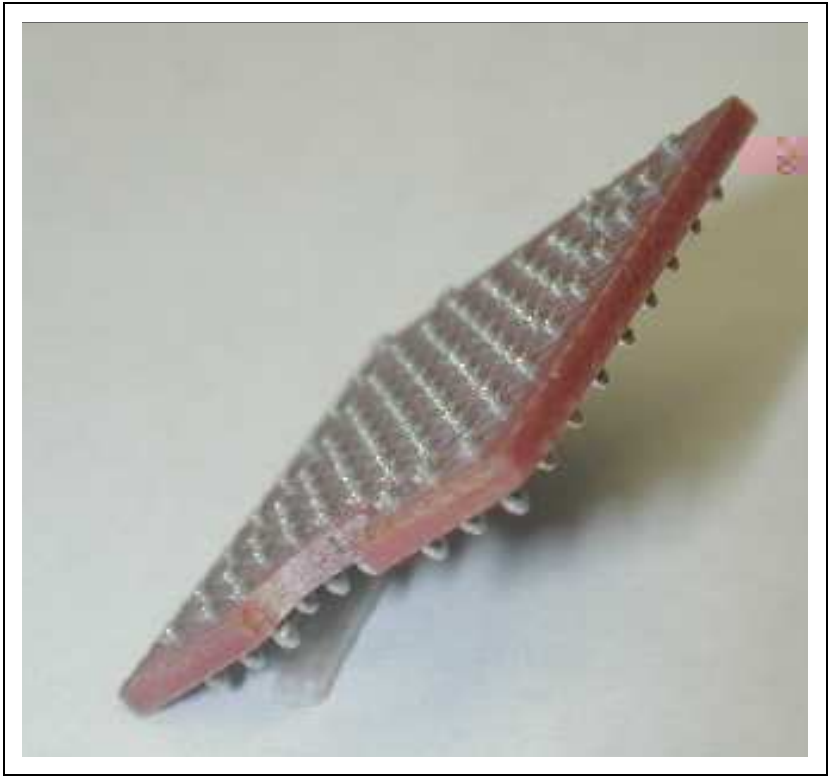}}
  \caption{Prototype of the surface-mount socket for the MAPMT.}
  \label{fi:Precicontact}
 \end{center}
\end{figure}

If one wants to avoid direct soldering of the MAPMT one needs a
suitable socket. A dedicated surface-mount socket 
was produced~\cite{bi:Precicontact}, as shown in
figure~\ref{fi:Precicontact}. 

\subsection{The Elementary-Cell assembly}

A possible assembly procedure can be carried on via the following steps.
\begin{enumerate}
\item
Mounting of the passive components on the Base-Board: 
voltage-dividers components, surface-mount socket for the MAPMT and connectors.
Test of the passive components. An X-ray scan can be performed at this stage to check the soldering joints.
\item
Installation and test of the front-end electronics.
\item
Installation and test of the MAPMT.
\item
Potting of the whole EC.
\end{enumerate}

A suitable potting between the MAPMT and the Base-Board and all around the four MAPMT 
with a suitable resin will ensure 
electrical and mechanical insulation, 
mechanical damping, 
structural strength, 
containment and good protection of the components, 
long-lasting mechanical resistance and good electrical contacts, 
and, possibly, light tightness and good thermal conduction.
Dow Corning DC-93500, a commonly used potting resin for space application, 
has been assumed as base-line potting compound.  

The impossibility to perform a complete visual inspection of the EC can be
overcome by X-ray inspection techniques and/or by defining a suitable
alternative testing functional procedure, either for the MAPMT
connections and for the front-end ASIC.

\begin{figure}[htb]
 \begin{center}
  \fbox{\includegraphics[width=0.99\textwidth]{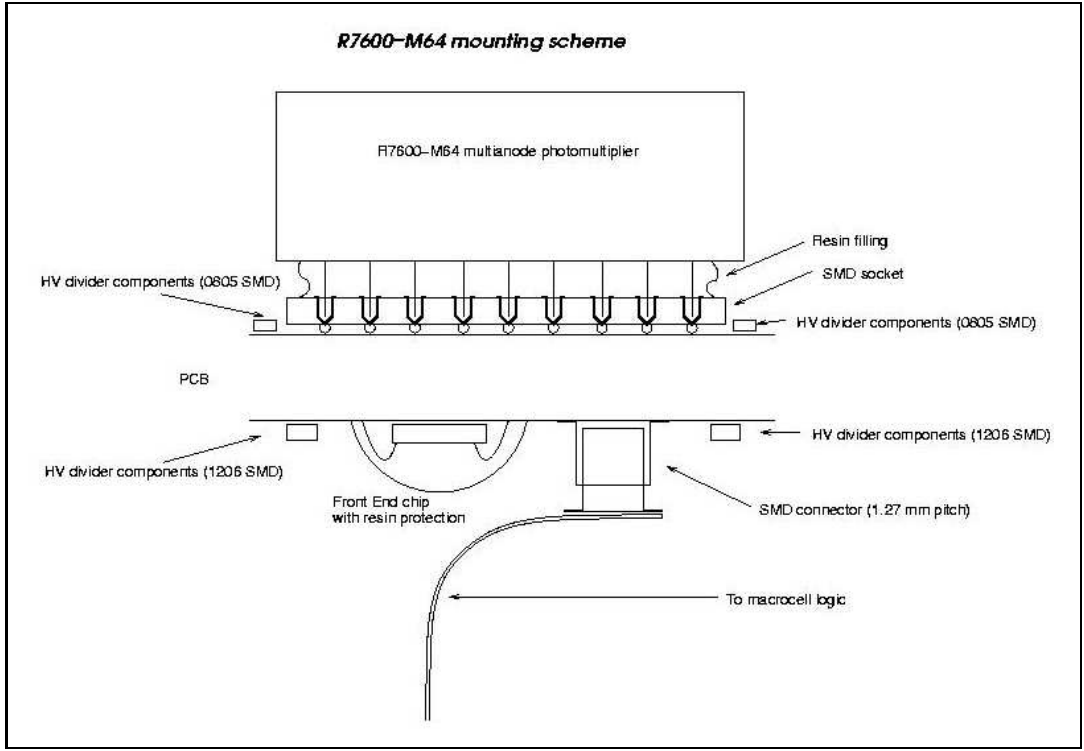}}
  \caption{Scheme of the mounting scheme for the EC (cross section).}
  \label{fi:mounting}
 \end{center}
\end{figure}

The scheme of the mounting is shown in figure~\ref{fi:mounting}.

\section{Prototype tests}
\label{pa:tests}

A batch of Base-Board prototypes has been manufactured to be used in laboratory tests, and equipped
with BGA sockets and dummy, expendable, MAPMT. This prototype is reasonably close to
the designed EC as far as the thermo-mechanical properties are concerned.
Moreover one prototype was installed including real MAPMT.

\begin{figure}[htb]
 \begin{center}
  \fbox{\includegraphics[width=0.99\textwidth]{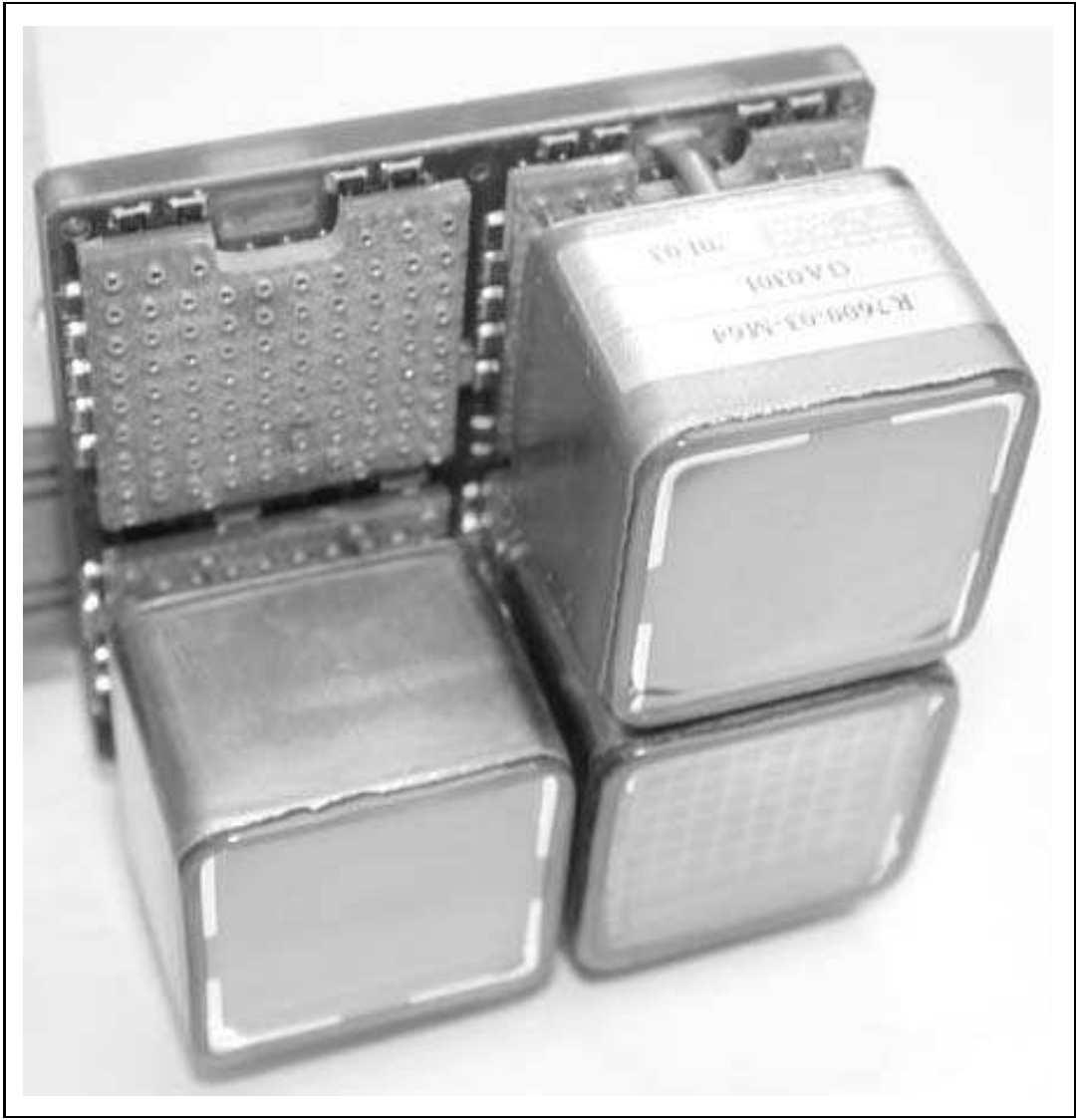}}
  \caption{Front-view of the prototype EC, 
    with three of the MAPMT installed but without Light Collector Systems and no potting.}
  \label{fi:fig1}
 \end{center}
\end{figure}

\begin{figure}[htb]
 \begin{center}
  \fbox{\includegraphics[width=0.99\textwidth]{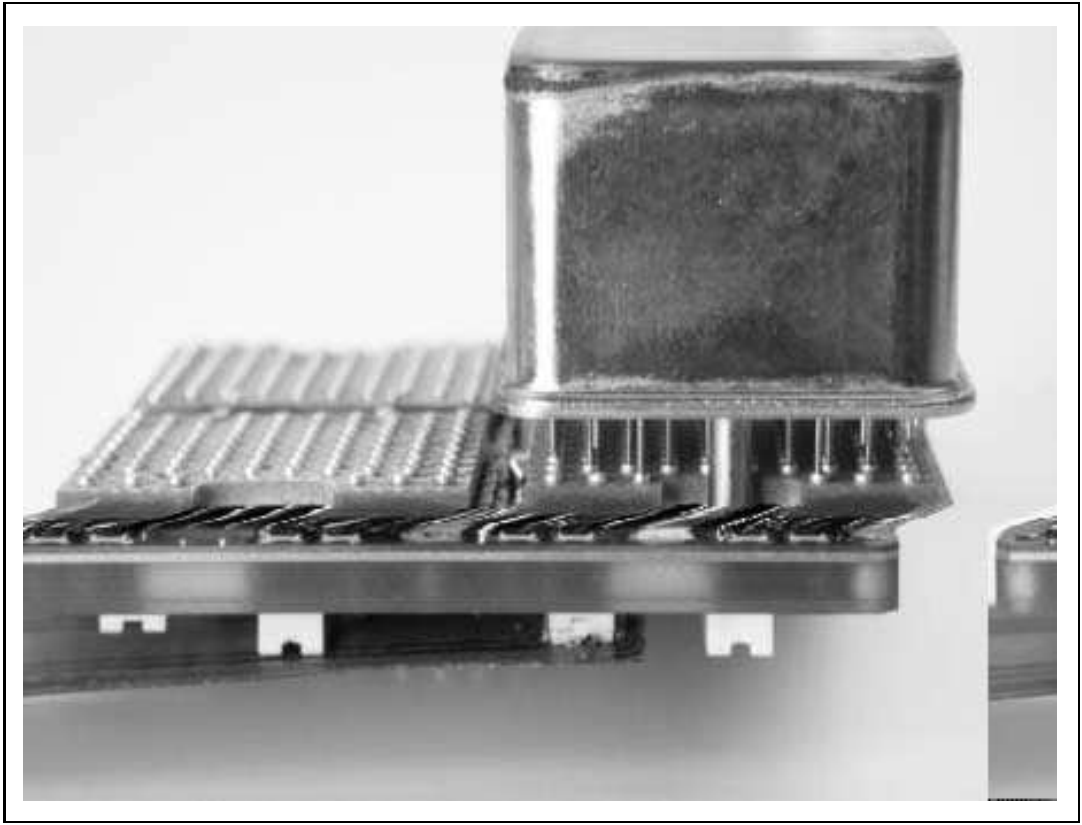}}
  \caption{Side-view of the prototype EC, 
    with three of the MAPMT installed but without Light Collector Systems and no potting.}
  \label{fi:fig3}
 \end{center}
\end{figure}

Pictures of the prototype EC with some of the MAPMT installed, without Light Collector Systems and
no potting, are shown in Figure~\ref{fi:fig1} and~\ref{fi:fig3}.

\subsection{Functional tests}

The EC prototype was submitted to functional tests. The behaviour of the MAPMT was as expected and
no effect due to the specific housing on the EC was noticed.

\subsection{Structural tests}

Effectiveness and reliability of $M2.5$ screws, a forced choice because of the tight layout
constraints, have been checked through an extensive a test campaign. An $M2.5$ stainless steel screw
has been tightened on a $5 \um{mm}$ thick FR4 board and submitted to approximately six times the
maximum expected equivalent static stress. After eight months, there was no evidence of damage
either of the Base-Board, or of the screw. Then, $M2.5$ stainless steel screws have been tightened
to rupture onto a $4.5 \um{mm}$ thick G10 board (including a $0.5 \um{mm}$ thick copper layer
inside): all the five screws tested broke at a torque in the range from $ (60 \div 65) \um{cNm}$.

Mechanical tests have then been performed on a vibrating table.

A EC sub-assembly, complete with dummy MAPMT and potting resin, has been set up. 
Its measured mass, including stainless steel screws and aluminium spacers,
was measured to be $0.16 \um{kg}$, that is 
much lighter than assumed in the simulations. 

The EC assembly has been installed, in
horizontal position, equipped with an accelerometer, and submitted to a \textit{sine sweep}.
The first resonances were found at $1.0 \um{kHz}$ and $1.3 \um{kHz}$. 
Afterwards a Random Vibration Test was started,
submitting the assembly to \textit{white noise} in the frequency range from $20 \um{Hz} \div 2000 \um{Hz}$.
The load was gradually stepped,
from $4 \um{G}$ RMS to $9 \um{G}$, and each step lasted 60 seconds. As a last step, the spectrum measured by the
accelerometer on the support bracket (input) was equivalent to  $8.3 \um{G}$. 
After the Random Vibration Test campaign, a
sine sweep has been performed again, and resonances have been found at almost the same frequencies,
showing that no significant changes happened. 

The same test campaign has been repeated with the EC assembly
installed in vertical position.
The first resonances were found at $0.6 \um{Hz}$ and $0.8 \um{Hz}$. 
The system has been submitted to the same Random Vibration Test with \textit{white noise} and 
up to $9 \um{G}$ in input was measured. 
The location of the resonances showed no significant change after the test.

At the end, the assembly in horizontal position has been submitted
to a \textit{sine burst}, where a sinusoidal excitation at $35 \um{Hz}$ and amplitude equivalent to 
$11 \um{G}$
entered the system for 10 seconds. The sine sweep again found almost the same resonances after the
test. The EC response was about the same as the input, showing a rather ineffective damping
at $35 \um{Hz}$. 

After all the tests, the whole assembly has been inspected, and electrical continuity
checked. No damage has been noticed.

The preliminary tests performed on the prototypes proved that the mechanical design solutions chosen are 
reliable, including the choice of the stainless steel screws holding
directly on a thread machined on the glass-epoxy EC Base-Board.

\subsection{Thermal tests}

A group of three EC, equipped with heaters to simulate the power dissipation of the
on-board electronics and instrumented with PT100 temperature probes, has been installed in a test
set-up, with a geometry that kept at negligible level the convective heat exchange.

All the parameters have been set to reproduce the boundary and load conditions, as assumed in the
simulations. Measured Base-Boards surface temperatures were around $22 \Cel$, slightly above room
temperature, while the temperature drop measured across the mechanical links of the EC
was in the range $4.5 \Cel \div 5.8 \Cel$. Results are in good agreement with the simulations. 
The heat balance
confirmed that $0.95$ of the power was flowing through the EC mechanical fasteners, as desired.

The good agreement between estimates and measurements confirms the good thermal conduction 
at the metallized holes.

The preliminary tests performed on the prototypes proved that the thermal design solutions chosen are 
reliable, including the choice of including a copper layer inside the EC Base-Board 
to reduce
both temperature peaks and temperature non-uniformities.

\section{Conclusions}

The design, construction and tests of the prototype
of a Base-Board for arrays of closely-packed Multi-Anode Photo-Multipliers was presented.
The design was carried on taking into account the tight requirements of space applications and
it is, therefore, of wide applicability.

The EC Base-Board design proved to be robust enough to withstand the severe environmental
conditions expected during a space mission.

The next and final step of prototyping foresees a Base-Board with the front-end electronics ASIC
installed on the back-side.
It is planned to use the device as the sensitive element of the Photo-Detector for
applications including: measurement of the background to Ultra-High-Energy Cosmic Rays observations
from space, test measurements of Ultra-High-Energy Cosmic Rays either from the Earth surface or
from balloons and measurements of the air scintillation yield (a basic information for
Ultra-High-Energy Cosmic Rays experiments).

\section{Acknowledgements}

The authors wish to thank Laura Confalonieri (Hamamatsu Italia) and
Yuji Yoshizawa (Hamamatsu Photonics Electron Tube Center)
for many useful discussions and support.

The authors are grateful to INFN for continuing support to the project.

\listoffigures

\end{document}